\documentclass[journal]{IEEEtran}

\usepackage{multirow}     
\usepackage{booktabs}     
\usepackage{amsmath,amssymb,amsfonts}
\usepackage{rotating}
\usepackage{graphicx}
\usepackage{xcolor}
\usepackage{algorithm}
\usepackage{algorithmic}
\usepackage{wrapfig}
\usepackage{caption}
\usepackage{subcaption}
\usepackage{adjustbox}
\usepackage{cite}

\begin{document}
\bstctlcite{IEEEexample:BSTcontrol}
\title{Sparse Compressed Spiking Neural Network Accelerator for Object Detection\\
}

\author{\IEEEauthorblockN{Hong-Han Lien and Tian-Sheuan Chang, \textit{Senior Member, IEEE}}
\thanks{This work was supported by the Ministry of Science and Technology, Taiwan, under Grant 110-2634-F-009 -017, 109-2639-E-009-001, 110-2221-E-A49 -148 -MY3. The authors are with the Institute of Electronics, National Yang Ming Chiao Tung University, Hsinchu 30010, Taiwan (e-mail: scott860228.ee08@nycu.edu.tw, tschang@nycu.edu.tw) }%
\thanks{
© 2021 IEEE.  Personal use of this material is permitted.  Permission from IEEE must be obtained for all other uses, in any current or future media, including reprinting/republishing this material for advertising or promotional purposes, creating new collective works, for resale or redistribution to servers or lists, or reuse of any copyrighted component of this work in other works.\\
H. -H. Lien and T. -S. Chang, "Sparse Compressed Spiking Neural Network Accelerator for Object Detection," in IEEE Transactions on Circuits and Systems I: Regular Papers, vol. 69, no. 5, pp. 2060-2069, May 2022, doi: 10.1109/TCSI.2022.3149006.
}
\thanks{Manuscript received XXXX, 2021; revised XXXX, 2021.}
}


\maketitle



\begin{abstract}%
Spiking neural networks (SNNs), which are inspired by the human brain, have recently gained popularity due to their relatively simple and low-power hardware for transmitting binary spikes and highly sparse activation maps. However, because SNNs contain extra time dimension information, the SNN accelerator will require more buffers and take longer to infer, especially for the more difficult high-resolution object detection task. As a result, this paper proposes a sparse compressed spiking neural network accelerator that takes advantage of the high sparsity of activation maps and weights by utilizing the proposed gated one-to-all product for low power and highly parallel model execution. The experimental result of the neural network shows 71.5$\%$ mAP with mixed (1,3) time steps on the IVS 3cls dataset. The accelerator  with the TSMC 28nm CMOS process can achieve 1024$\times$576@29 frames per second processing when running at 500MHz with 35.88TOPS/W energy efficiency and 1.05mJ energy consumption per frame.


\end{abstract}









\section{Introduction}

Spiking neural networks (SNNs) have gotten a lot of attention as an alternative to artificial neural networks (ANNs) because the event-driven processing paradigm, which aims to mimic the brain, can achieve competitive accuracy while using very little power. Furthermore, the binary spike transmission is relatively simple to implement in hardware. SNNs, as opposed to ANNs, use spike input and output as well as continuous-time leaky integrate-and-fire (LIF) neurons. We use a discrete-time approximate LIF with a delta-shaped synaptic kernel in this paper. The input spike will be multiplied by weights and accumulated with the residual potential from the previous time step at each time step. If the potential exceeds the threshold, the output spike will be fired. Thus, with high sparsity and simple spike I/O, its hardware implementation has got popular in recent years for low power and real-time model execution.


Various accelerators have recently been proposed, which can be divided into two approaches based on their supported topologies~\cite{khodamoradi2021s2n2}: general mesh and feedforward. For general mesh topology, large-scale neuromorphic hardware systems such as TrueNorth~\cite{akopyan2015truenorth}, Loihi~\cite{davies2018loihi}, SpiNNaker\cite{painkras2013spinnaker}, BrainScaleS\cite{Brainscales}, ODIN\cite{ODIN}, $\mu$Brain\cite{mBrain} and DYNAPs\cite{DYNAPs} support a mesh of neurons with no particular topology by advanced routers and schedulers. They distribute the SNN network among the neurocores and use many small neurocores for computations with specially arranged memory and processing elements (PEs). Each neurocore is responsible for storing a portion of the weights and computing that portion of the SNN topology. 
They are biologically plausible but need large area costs. For the feedforward topology, designs such as \cite{khodamoradi2021s2n2, narayanan2020spinalflow, chen202167, park20197, yin2017algorithm} processes a feedforward network in a systolic fashion. Their neurocores are arranged either in a cascaded fashion\cite{khodamoradi2021s2n2, chen202167, park20197, yin2017algorithm} or in a configurable PE array\cite{narayanan2020spinalflow}.  Our work also falls into this category and is similar to Spinalflow's approach\cite{narayanan2020spinalflow}. 

However, in contrast to ANN accelerators, current SNN architectures achieve lower throughput and higher energy per neuron, according to ~\cite{narayanan2020spinalflow,narayanan2017inxs}. This phenomenon is primarily due to the extra time dimension information in SNNs\cite{narayanan2020spinalflow,chen202167}, which causes the longer inference time and requires to repeatedly access the same weight data from off-chip memory and thus consumes a lot of unnecessary energy. One way to solve this problem is to use a large internal memory to fit all weights in one layer, but this occupies a lot of chip area. Another approach is to use binary weights to overcome this problem~\cite{lien2021vsa}. However, when applied to more difficult tasks, it is easy to cause a significant drop in accuracy, which is not a good solution.

In addition, the extra time dimension information in SNNs also makes it difficult for the accelerator to execute the network in real time. Therefore, using the sparsity of the weights to lower the area and speed up the inference is a potential method. SpinalFlow~\cite{narayanan2020spinalflow} proposed an architecture
that can skip the zero activation computation. However, their sequential processing method sacrifices massive parallelism and only supports the temporal coding inputs. Furthermore, they ignore the sparsity of the weights, resulting in a filter buffer size of up to 576 KB and a low area efficiency. Sparsity designs have also been explored in ANNs to significantly speed up the inference through the zero-skipping control. SCNN~\cite{parashar2017scnn} uses the Cartesian product to multiply all nonzero inputs and all nonzero weights, and then accumulates the partial product of the output of the corresponding coordinate with the help of the scatter network.
However, the complex scatter network dramatically increases the extra overhead of hardware. SparTen~\cite{gondimalla2019sparten} uses the space-efficient bit-mask representation to represent the sparse data and proposes sparse vector-vector multiplication
that finds the matching nonzero bit positions by simple logic to accumulate the corresponding partial product. In summary, existing works need complex control to deal with the sparse nature of activation maps, which causes a lot of extra power consumption or sacrifices parallelism to achieve real-time inference.

This paper proposes a sparse compressed SNN accelerator with spatial parallelism for real-time object detection to address the aforementioned issues. The energy efficient object detection model is designed with hardware-friendly low complexity modules. Its model is compressed and encoded with the bit mask format for smaller model size and sparse weights to reduce the on-chip memory requirement. Furthermore, to exploit sparsity without complex control issues, we propose a gated one-to-all product that efficiently utilizes the sparse feature with high parallelism and low power. The hardware design can be reconfigured for SNN layers as well as input encoding layers. Besides, this hardware adopts the spatial parallelism to better support the large input resolution requirement of object detection. The final implementation can attain 1024$\times$576@29 frames per second processing when running at 500MHz with 35.88TOPS/W energy efficiency and 1.05mJ energy consumption per frame.

The rest of the paper is organized as follows. Section II shows the proposed energy efficient model. Section III presents the hardware architecture. Section IV shows the experimental results and comparisons. Finally, this paper is concluded in Section V.







\section{Energy Efficient Model Design}
\subsection{Network Architecture}
The SNN model design is mainly divided into three training methods: (a) unsupervised training, (b) ANNS-to-SNNs conversion, and (c) direct training with backpropagation. The most commonly used algorithm for unsupervised training is spike-timing-dependent plasticity (STDP)~\cite{bi1998synaptic}. This local training method avoids a large amount of gradient storage and is very suitable for on-chip learning. However, it lacks global information, which makes it difficult for large-scale networks and advanced tasks. For the object detection task, Spiking-YOLO~\cite{kim2020spiking} is the first work that applies the deep SNNs to object detection. Their proposed channel-wise normalization enables a higher firing rate, which leads to fast information transmission and obtains better results with fewer time steps, but still requires up to one thousand time steps due to its ANNs-to-SNNs conversion. Recent ANNs-to-SNNs conversion methods have improved the latency to a few time steps by the hybrid training\cite{rathi2020diet, datta2021can} (ANN pre-training followed by SNN training with tunable threshold and leakage).
To achieve accurate and real-time inference by hardware accelerators, we use the learning-based spatio-temporal backpropagation (STBP)~\cite{wu2019direct} with threshold-dependent batch normalization (tdBN)~\cite{zheng2020going} to train our object detection network that only takes a few time steps for inference. 

\begin{figure}[htb]
\centering
\includegraphics[height=!,width=0.4\linewidth,keepaspectratio=true]{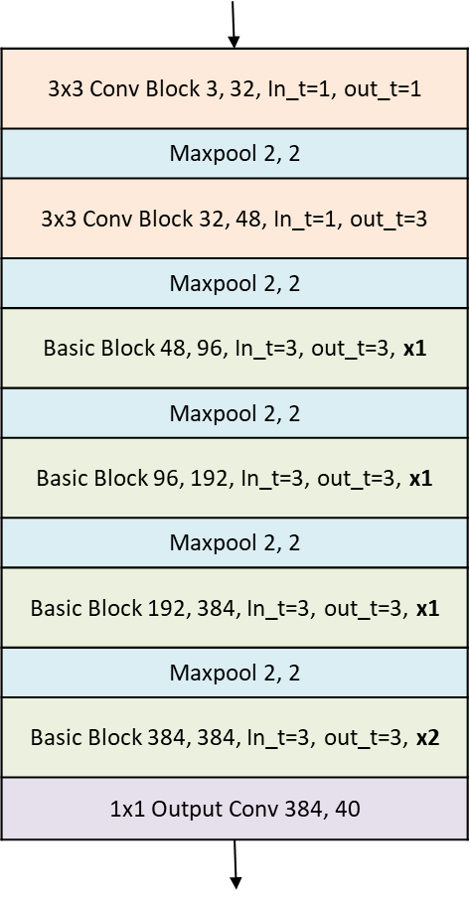}
\caption{The network architecture for object detection, where in\_T and out\_T are time steps for input and output, respectively. }
\label{od model fig}
\end{figure}

\begin{figure}[htb]
\centering
\includegraphics[height=!,width=0.9\linewidth,keepaspectratio=true]{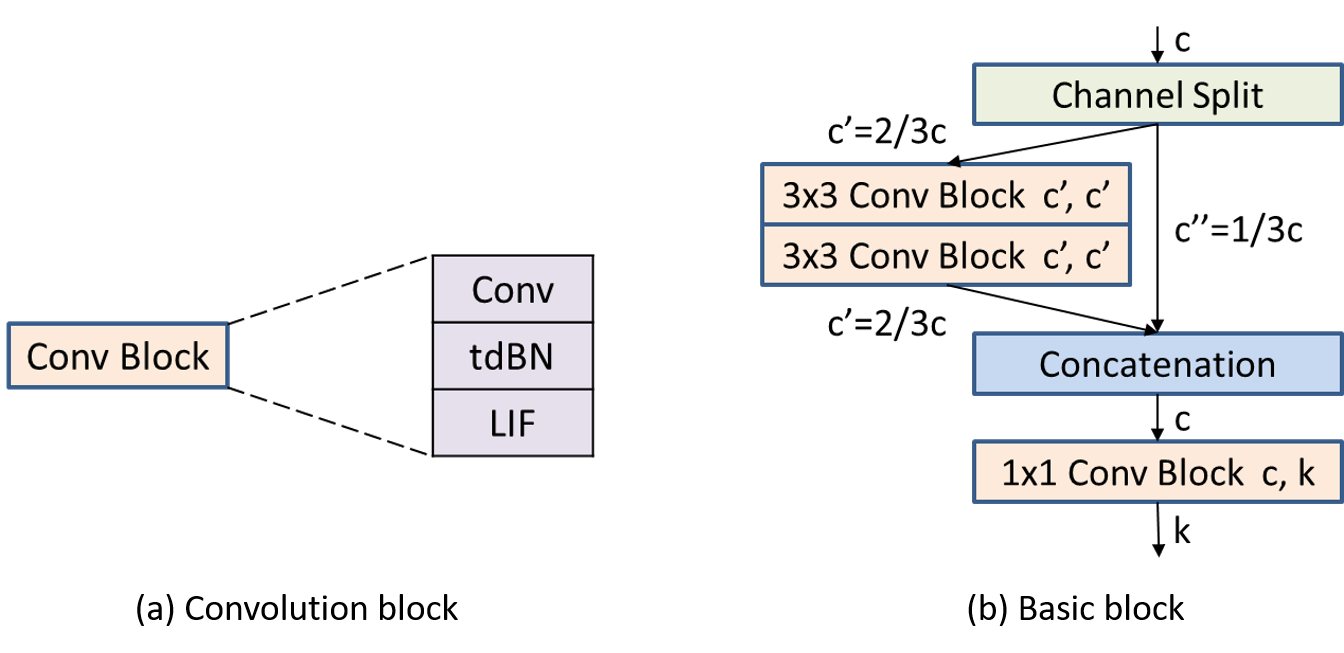}
\caption{(a) Convolution block (b) Basic block}
\label{csp basic block fig}
\end{figure}

Fig.~\ref{od model fig} shows the proposed SNN network for object detection with the RGB input images. The input layer uses the encoding convolution block to encode the multibit inputs into spikes for the following SNN layers. Then following the input layers, we adopt the convolution block and several basic blocks as shown in Fig.~\ref{csp basic block fig}.  The convolution block as in  Fig.~\ref{csp basic block fig}(a) use tdBN for low inference time steps. The basic block as in Fig.~\ref{csp basic block fig}(b) adopts the CSPNet~\cite{wang2020cspnet} module, which reduces the number of operations through channel splitting and then uses 1$\times$1 convolution to aggregate the features. The structure of the basic block not only speeds up the inference but also helps the gradient propagation. The number of channels for the shortcut connection is set to half of the number of channels for the stacked convolution path to reduce the parameters and operations in 1$\times$1 convolutional layers. The final Output Convolution layer accumulates the membrane potential with no reset and averages the output of all time steps. This layer is for the final object detection head, which adopts the detection method of YOLOv2~\cite{redmon2017yolo9000}. In this model, the threshold of LIF is set to 0.5, and the leaky term of LIF is set to 0.25 for a simple hardware implementation.

Our SNN model uses mixed time steps for the trade-off between computation latency and accuracy. As shown in Fig.~\ref{od model fig}, the input time step is set to one for the first two convolutional layers and three for others.
The first convolutional layer is treated as an ANN layer for input encoding that fires once based on the current input. The second convolutional layer receives one-time-step input and generates three-time-step outputs to the following layer. This layer computes the convolution part once and passes the same output to the LIF for three time steps to produce three different outputs. The other layers use three time steps.
The time step of the first two layers is set to one to reduce the number of operations significantly due to their large input. Such method only has a small impact on the final regression result because the first few layers are mainly to extract the features of the contour.

\subsection{Block Convolution}
For ease of processing, input feature maps are frequently partitioned into tiles. However, due to overlapped processing, the partial sum at the tile boundary must be stored in buffers, which consumes a significant amount of area. To avoid this problem, we adopt the block convolution method~\cite{li2021block} that partitions the input into non-overlapped blocks and computes each block independently without data dependency, saving large amounts of boundary buffers or access energy.  We set the block size to 32$\times$18 in this paper with replicate boundary padding.




\subsection{Fine-grained Pruning}

\begin{figure}[htb]
\centering
\includegraphics[height=!,width=1.0\linewidth,keepaspectratio=true]{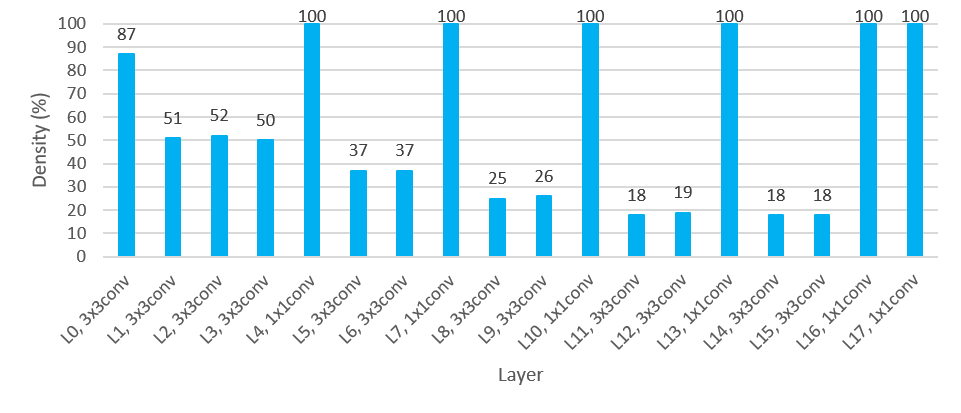}
\caption{The density of pruned weights of each layer}
\label{pruned weights fig}
\end{figure}

To significantly reduce the weight memory requirements and speed up inference, we apply the fine-grained pruning~\cite{han2015learning} to this network. This method sets weights below the threshold to zero, where the threshold is determined by the pruning rate. In this paper, we set the pruning rate to 80$\%$ to prune 3$\times$3 kernels and keep all weights of the 1$\times$1 kernels intact. The final pruned result is shown in Fig.~\ref{pruned weights fig}. The pruned model reduces 70$\%$ of weights and 47.3\% of operation counts, respectively. Note that there are several SNN dedicated pruning methods that could be applied as well to further improve pruning rate. For example, \cite{SNNadmm} combines STBP and alternating direction method of multipliers (ADMM) to prune models via direct training, which needs iterative tuning and training for optimum performance.
 The pruning during training approach uses output firing characteristics\cite{shi2019soft}, STDP based pruning\cite{Rathipruning} or gradient rewiring\cite{chen2021pruning} for joint learning of connectivity and network weight. Pruning based on the ANN-to-SNN conversion prunes the ANN model before conversion with attention guide\cite{kundu2021spike}. 

\subsection{Mixed Time Steps}
Although we used fine-grained pruning to speed up the inference, the speedup has not yet reached our expectation. The main reason is that the feature size of the first few layers is too large and the pruning tends to retain more weights of the first few layers as shown in Fig.~\ref{pruned weights fig}. Thus, the overall operation counts are still large. Therefore, we further reduce the number of operations by the proposed mixed time steps. 

\begin{figure}[htb]
\centering
\includegraphics[height=!,width=1.0\linewidth,keepaspectratio=true]{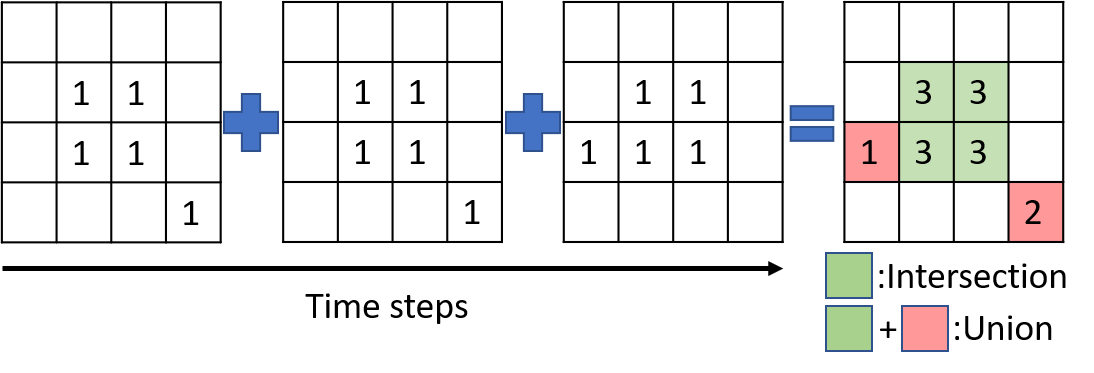}
\caption{An example of the mIoUT metric}
\label{mIoUT fig}
\end{figure}

To measure the similarity of features at different time steps, we propose a metric named mean Intersection over Union across Time-steps (mIoUT) as (\ref{eq miouat}).
\begin{equation}
\label{eq miouat}
    mIoUT=\frac{1}{C}\sum_{c=1}^{C}\frac{Number\; of\; Intersection_{c}}{Number\; of\; Union_{c}}
\end{equation}
where C is the channel number, the Intersection is defined as the firing count of the neuron equal to the total time steps, and the Union is defined as the firing count of the neuron  greater than zero but smaller than the total time steps. Fig.~\ref{mIoUT fig} shows an example of the metric, which accumulates spikes across time steps. The final result shows that there are four neurons fired at each time step and two neurons fired fewer than three times but greater than zero. Thus, the mIoUT is 0.67, which means the features of each time step are very similar.

\begin{figure}[htb]
\centering
\includegraphics[height=!,width=1.0\linewidth,keepaspectratio=true]{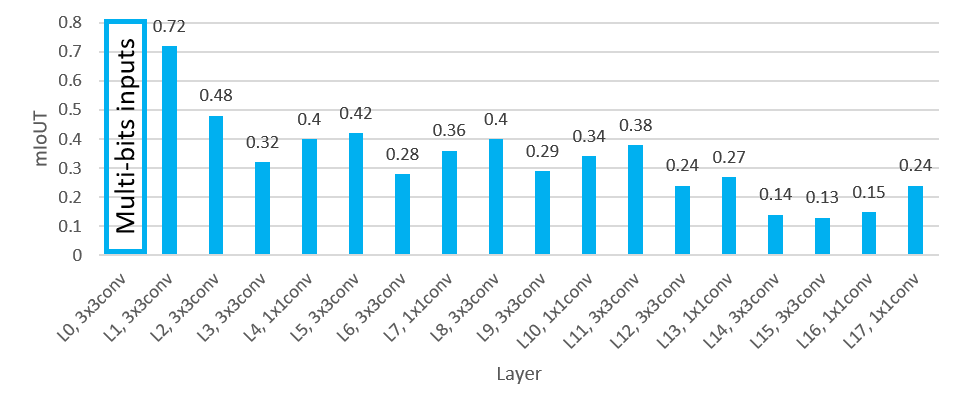}
\caption{mIoUT of input features at each layer}
\label{mIoUT of model fig}
\end{figure}

Fig.~\ref{mIoUT of model fig} shows the evaluation of the model with three time steps. This figure shows that the features of each time step at the second layer are very similar. Therefore, the time step number of the second layer is reduced to one, and the same convolutional result is passed through the LIF to produce three different outputs to the next layer. The model with (1, 3) mixed time steps can reduce 4.13 giga operations, which is 17\% reduction when compared to the original model.

\section{System Architecture}
\subsection{Analysis of Design Parallelism}
  To fulfill the high computation demand due to the large input and model size, some form of parallelism is necessary. For sparse SNN computation, we evaluate three possible cases.
  
\begin{figure}[ht]
\begin{subfigure}{1.0\linewidth}
\centering
  \includegraphics[height=!,width=1.0\linewidth]{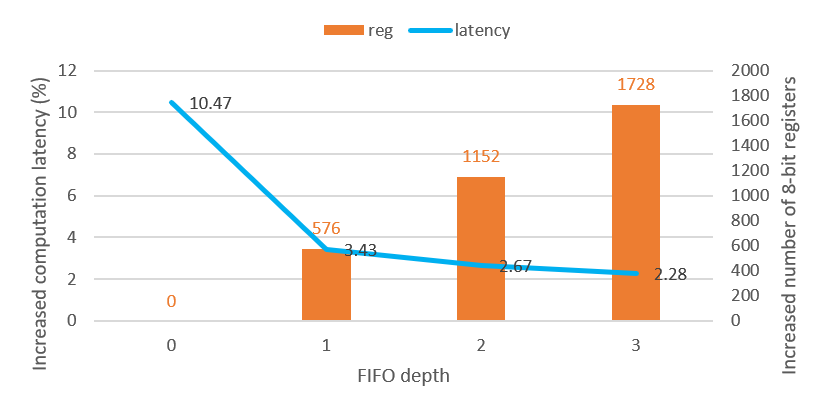} 
  \caption{Analysis of the input channel parallelism relative to the spatial parallelism}
  \label{Analysis of input channel parallelism fig}
\end{subfigure}
\begin{subfigure}{1.0\linewidth}
\centering
  \includegraphics[height=!,width=1.0\linewidth]{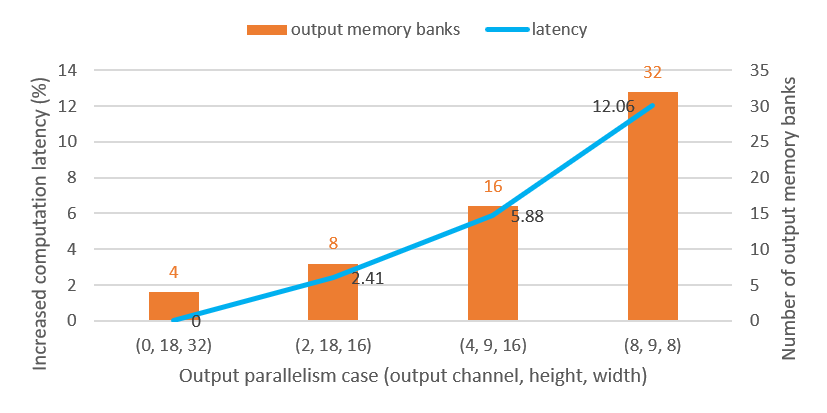}  
  \caption{Analysis of the output channel parallelism relative to the spatial parallelism}
  \label{Analysis of output channel parallelism fig}
\end{subfigure}
\caption{Analysis of design parallelism schemes}
\label{Analysis of design parallelism fig}
\end{figure}

\begin{enumerate}
\item Input channel parallelism: Because of the sparse compressed weight, if we have parallelism along the input channel, the execution speed of each channel will be different. Fig.~\ref{Analysis of design parallelism fig} (a) show the analysis of the input channel parallelism compared to the spatial parallelism. We allocate a total of 576 PEs used in this paper to process features on different dimensions. For the comparison purpose, we set the PE organizations as (input channel, height, width) = (8, 9, 8) to process 8 input channels with 9x8 feature size per channel. We also add FIFOs to reduce the latency due to the workload imbalance of input parallelism. In comparison to the spatial parallelism case ((0, 18, 32), denoted in the figure as the FIFO depth = 0 case), the input parallelism cases require more computation latency and more FIFO hardware resources as shown in the figure. The deep FIFOs will consume a large amount of chip area and power, potentially even larger than the area of PEs.
\item Output channel parallelism: If we have parallelism along the output channel, all output channels will use the same input feature. Thus, this design must wait until all output channel computations have been completed before moving on to the next input feature. Fig.~\ref{Analysis of design parallelism fig} (b) shows the simulation result compared to the spatial parallelism case (0, 18, 32). As shown in the figure, the computation latency is increased as more PEs are assigned to process features on the output channel dimension. Besides, to support output parallelism, this design needs to split SRAM buffers into more banks, necessitating the use of additional peripheral circuits.
\item Spatial parallelism: There will be no workload imbalance if we have parallelism along the spatial dimension. As a result, we don't need to connect any accumulators or buffers to the PEs. In addition,  according to \cite{li2021block}, the tile size of block convolution can be larger, which will have a smaller impact on detection result. The only restriction is that the input size be large enough to reap the benefits.
\end{enumerate}
Based on these observations, we finally choose the third case to reach the ultimate performance.

\begin{figure}[htb]
\centering
\includegraphics[height=!,width=1.0\linewidth,keepaspectratio=true]
{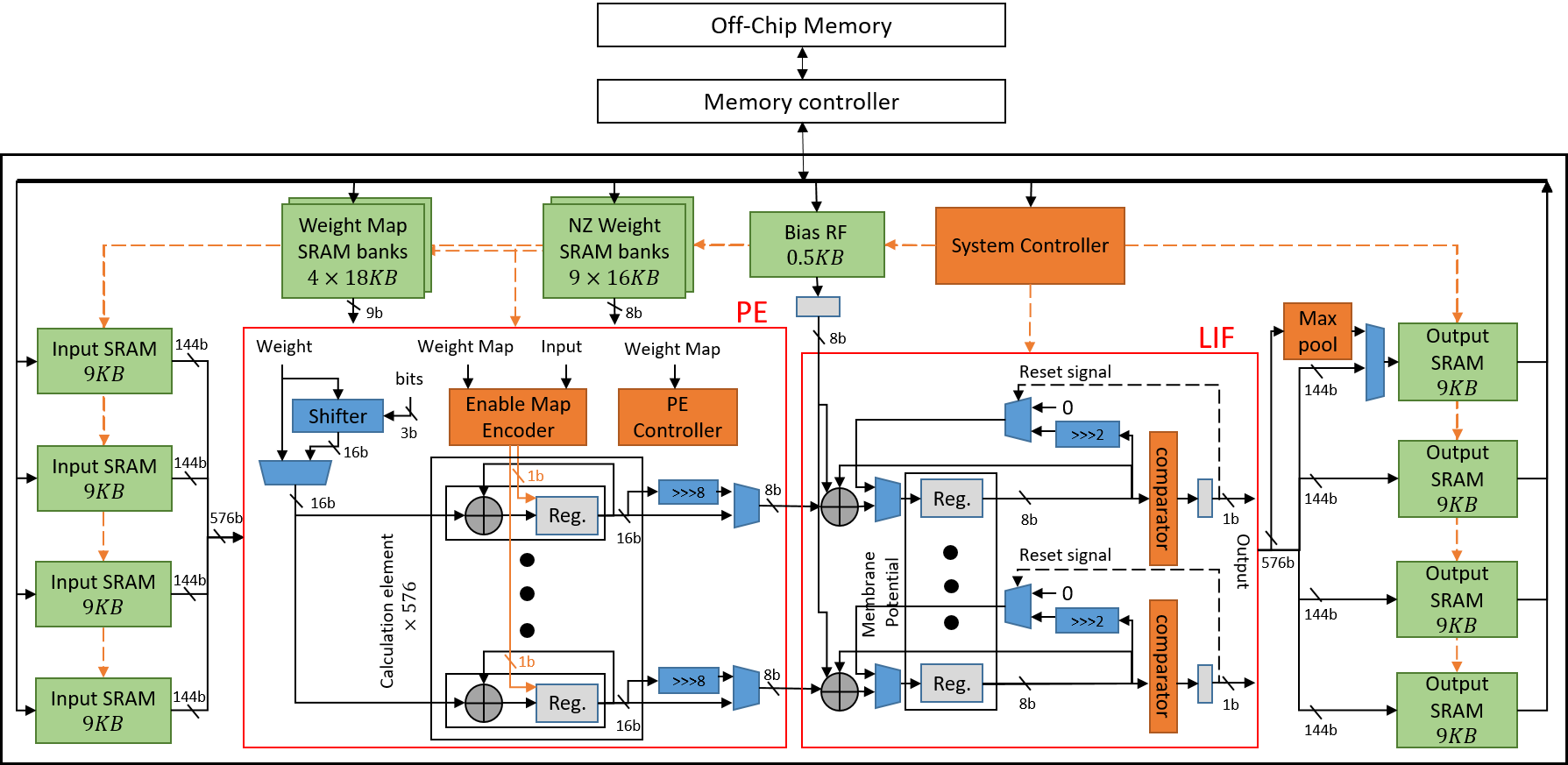}
\caption{The proposed system architecture for object detection}
\label{od system architecture fig}
\end{figure}

\subsection{Overview}
Fig.~\ref{od system architecture fig} shows the proposed system architecture, which consists of a PE module for sparse convolution computation, a LIF module for updating the membrane potential and firing spikes, a max-pooling module composed of simple OR gates, and a system controller module for controlling the memory and communication of each module based on configuration. The sparse compressed weight data is stored in the Weight Map SRAM bank and NZ Weight SRAM bank, respectively. Each of the four Input SRAM buffers stores a sub-tile of the activation map independently, just like the four Output SRAM buffers.

To reuse data, the proposed design will first access a tile of the input map and weights in the SRAM buffers. If the input is multibit, it will be split into bit planes, stored in SRAM, and then processed in the PE with the shifter and adder to complete the multibit convolution with bit serial input.

\subsubsection{Gated One-to-All Product}
Exploring the sparsity of SNN can help reduce computation and storage. However, because of the unexpected zero distribution, the processing of sparse activation maps is frequently irregular, reducing hardware parallelism and increasing control cost. Because of the simple spike input of SNN, this cost becomes significant. As a result, rather than using zero activation skipping to accelerate inference, this paper employs zero activation gating to reduce energy consumption, denoted as gated one-to-all product. This method avoids zero-weight computation to save cycle counts and multibit weight access and storage while maintaining high hardware parallelism for object detection computation.

\begin{figure}[htb]
\centering
\includegraphics[height=!,width=1.0\linewidth,keepaspectratio=true]
{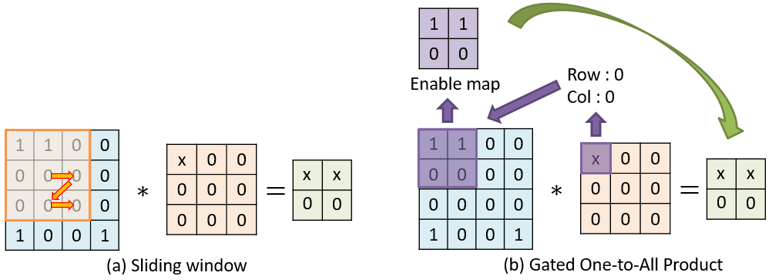}
\caption{The example of convolution with (a) sliding window, and (b) gated one-to-all product, where x denotes a weight value.}
\label{gated one-to-all product fig}
\end{figure}

Fig.~\ref{gated one-to-all product fig} shows an example of the proposed gated one-to-all product, which has a 4x4 input convoluted  with a 3x3 kernel to generate four outputs with a sliding window operation. For demonstration purpose, we assume only one nonzero weight in the kernel. 

\begin{figure}[htb]
\centering
\includegraphics[height=!,width=0.6\linewidth,keepaspectratio=true]
{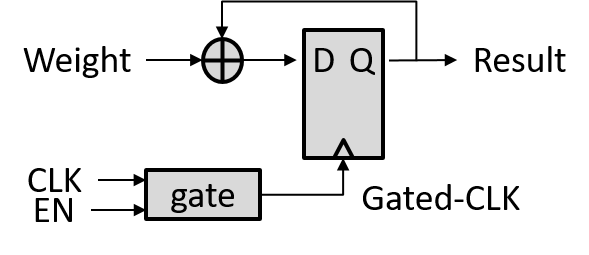}
\caption{The gated computation element in the PE}
\label{calculation element fig}
\end{figure}

For the proposed gated one-to-all product, we first encode an enable map according to each nonzero weight position in the kernel. In this example, the nonzero weight is at the upper left corner, $(row, col) = (0, 0)$. This weight will be multiplied with the four inputs at the upper left to generate output. The four input block is the enable map. These multiplications can be parallelly computed in hardware as a one-to-all product. If more nonzero weights exist in a kernel, the corresponding enable map will be shifted $R$ steps down and $C$ steps right for a nonzero weight at $(R, C)$ position in the kernel. These partial outputs will be accumulated into the final output. During this process, the zeros in the enable map will be used as a clock gating control to save power as shown in Fig.~\ref{calculation element fig}. This design introduces a gate module to replace the multiplication to reduce the dynamic power of each neuron. If the signal EN is 1, the clock will be enabled, and the corresponding weight will be accumulated. On the contrary, if the signal EN is 0, the clock will be turned off to keep the partial sum.  

In summary, the steps in the gated one-to-all product are: (1) find the enable map based on the nonzero weight position in a kernel, and (2) gate the output neurons with zero values in the enable map and accumulate the nonzero weight values of all enabled output neurons.



\subsubsection{Weight Compression Format}
\begin{figure}[htb]
\centering
\includegraphics[height=!,width=0.8\linewidth,keepaspectratio=true]
{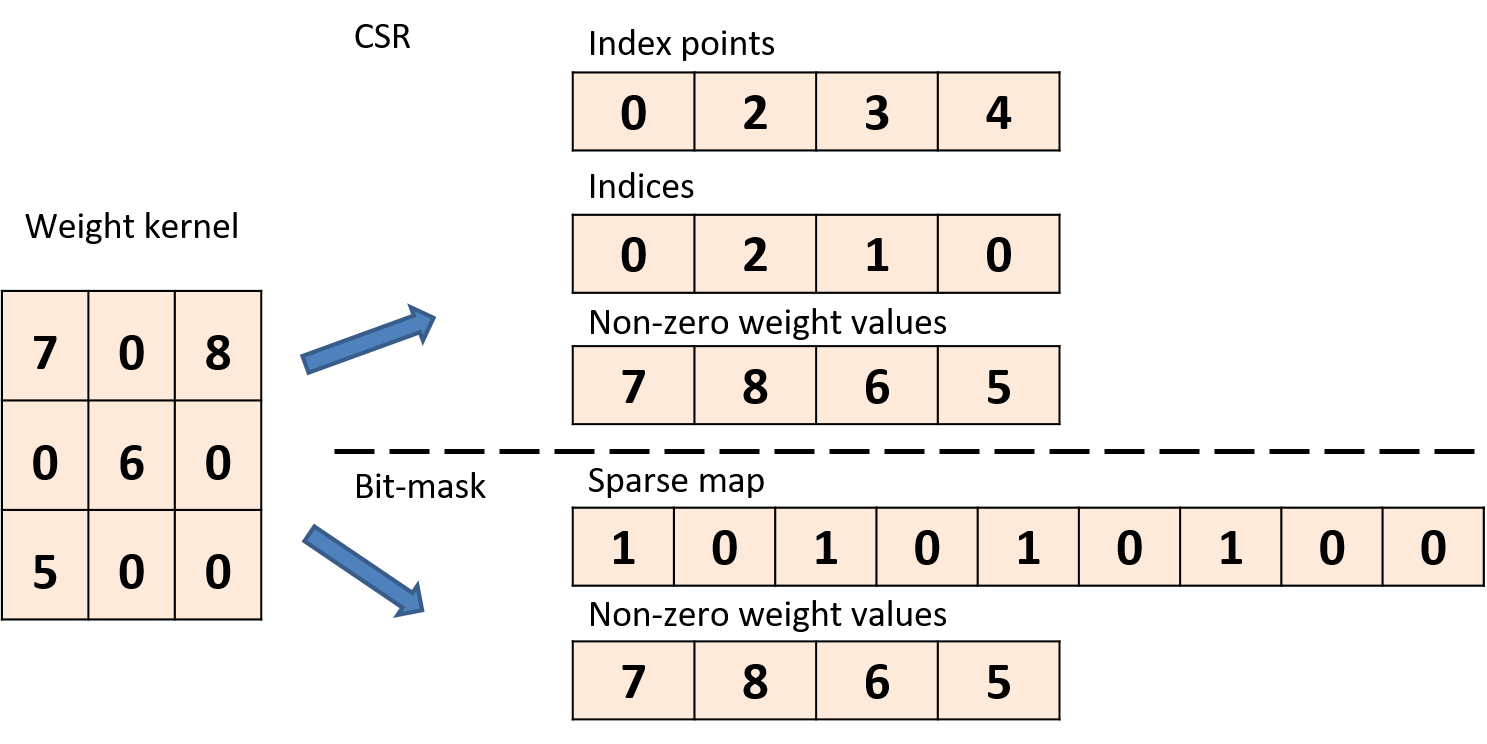}
\caption{Example of difference between CSR and bit-mask representation}
\label{csr & bit-mask fig}
\end{figure}

To support above gate one-to-all product, we choose the bit-mask representation to represent the sparse weight data instead of Compressed Sparse Row(CSR) used in other high-performance computing designs. Fig.~\ref{csr & bit-mask fig} illustrates the difference between CSR and bit-mask representation. CSR divides the kernel into three parts: index points, indexes, and non-zero weight values, while bit-mask representation divides the kernel into sparse map and non-zero weight values. According to the degree of sparsity of our network weights, we find that bit-mask representation can save more hardware cost due to its simplicity and is also more efficient for our sparse computation. 

\subsection{Data Flow}
\subsubsection{Example of the Proposed Convolution Data Flow}
\begin{figure*}[htb]
\centering
\adjustbox{trim={.0\width} {.55\height} {0.0\width} {.0\height},clip}%
  {\includegraphics[height=!,width=0.85\linewidth,keepaspectratio=true]
{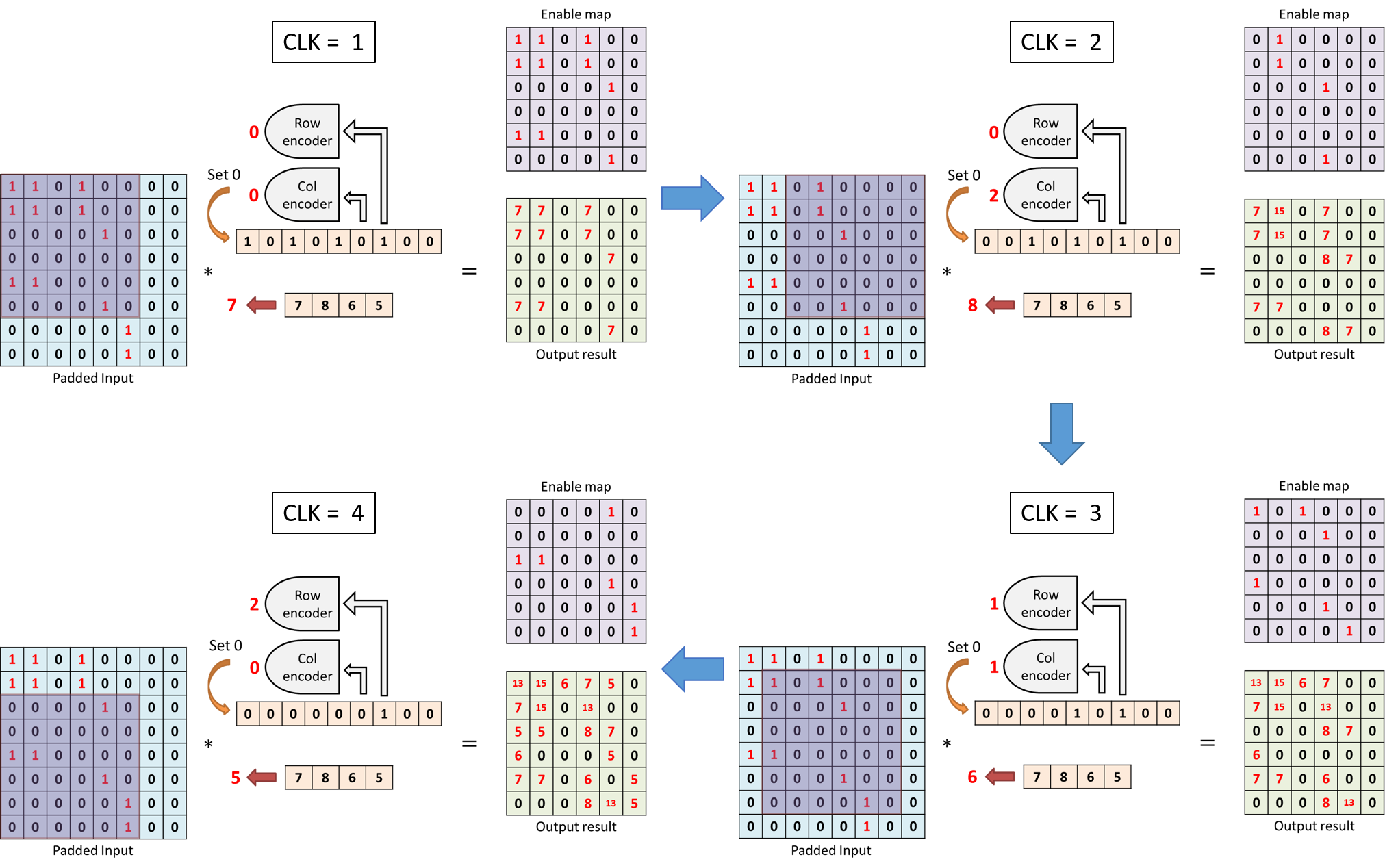}}
\caption{An example of the proposed convolution data flow.}
\label{data flow of convolution fig}
\end{figure*}

Based on the above gated one-to-all product, Fig.~\ref{data flow of convolution fig} shows an example of convolution data flow for the proposed architecture, where the input size is 8$\times$8 for clarity. The proposed PE will first load a channel of the input tile and the corresponding weight map and the nonzero weight values. The column encoder and row encoder will encode the position of the leftmost nonzero value of the weight map to find the enable map and compute the convolution by using the gated one-to-all product. 
Then, the leftmost nonzero value of the weight map will be cleared to zero before the next clock. At the second cycle, the PE will find the next nonzero weight and repeat the above process again. The output result of the second cycle will be accumulated with the one of the first cycle. This will be repeated until the final nonzero weight. During this process, the zero weight values will be skipped to reduce cycle counts.  


\subsubsection{Algorithm For The Proposed Computing Flow}
\begin{figure}[t]
\centering
\includegraphics[height=!,width=1.0\linewidth,keepaspectratio=true]
{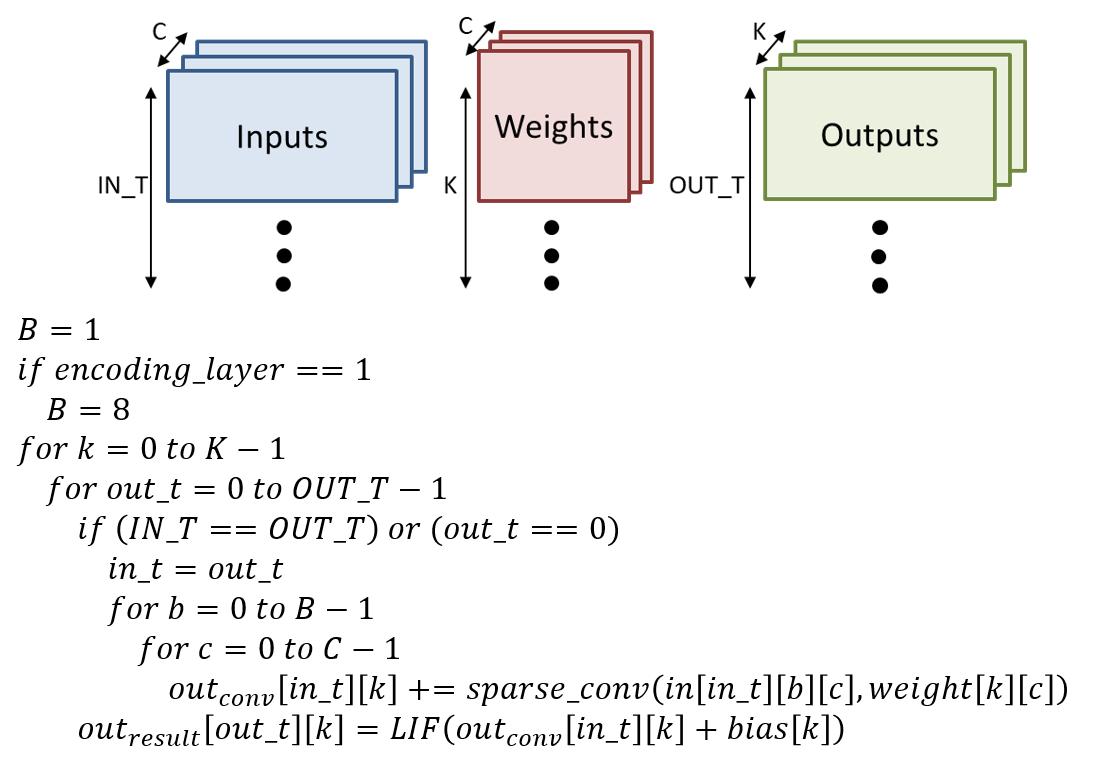}
\caption{Data flow of spiking neural network}
\label{data flow of snn fig}
\end{figure}

Fig.~\ref{data flow of snn fig} shows the proposed computing flow of our SNN design, which can be concisely described as the nested loop order: \textit{output channel} $K$ $\rightarrow$ \textit{time step} $T$ $\rightarrow$  \textit{input bit number} $B$ $\rightarrow$ \textit{input channel} $C$. 
The whole computing flow is described below. Before proceeding with main processing, we will configure the input based on the layer type and time step number. In our model, we have a normal SNN layer for spike input and an encoding layer for the multibit RGB input image. To provide unified support for both layers, the multibit input will be split into bit planes and processed bit serially as the spike input. Thus, we introduce the  dimension \textit{input bit number} $B$ and then the dimension \textit{input channel} $C$, and set B to eight for the input encoding layer. B is set to one for all other layers.  The second input configuration is based on the time step number $T$ since the input time step of the first two layers is set to one for lower operations and the other layers use a higher time step. If the input time step is not equal to the output time step, the convolutional result is repeated for each output time step, resulting in different outputs after the LIF operation. Following  the input configuration, the entire loop will be executed as shown below. For each output channel, each output time step, each bit plane, and each input channel, apply the sparse convolution as the gated one-to-all product, and output the final result after the LIF operation. 

\begin{figure}[t]
\centering
\includegraphics[height=!,width=1.0\linewidth,keepaspectratio=true]
{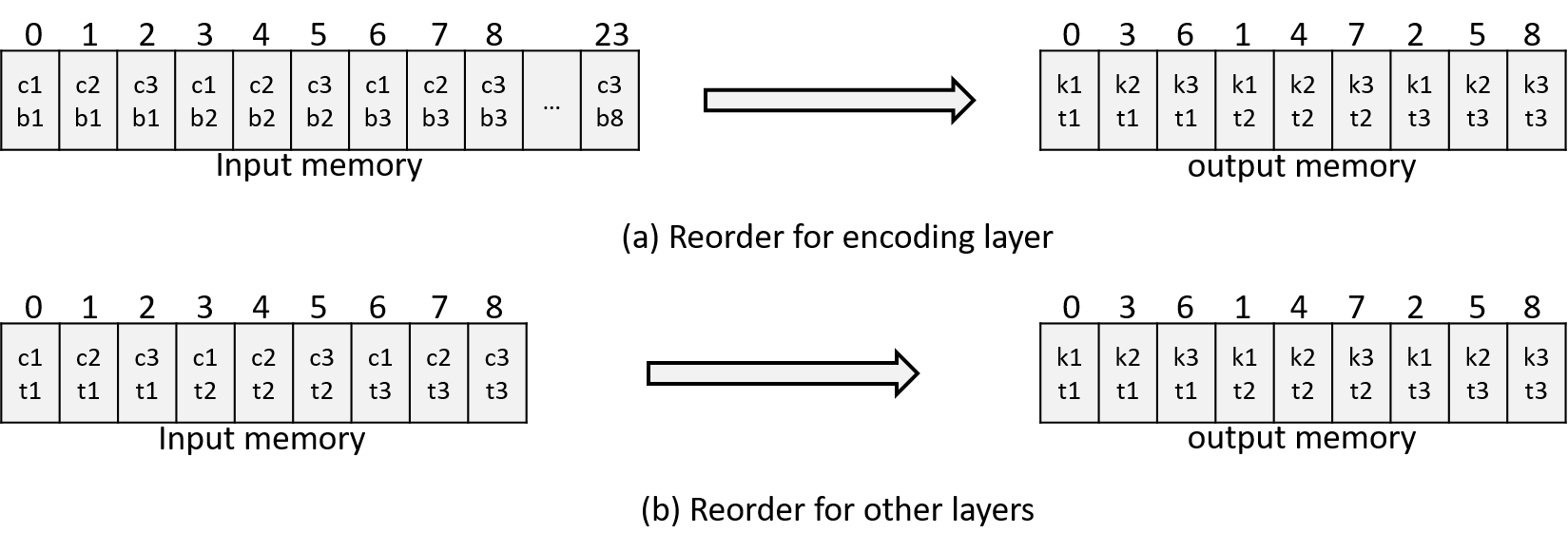}
\caption{(a) Reorder for encoding layer (b) Reorder for other layers}
\label{temporal channel reorder fig}
\end{figure}

In addition, according to the $KTBC$ loop order, a layer will complete the \textit{input channel} $C$ dimension  before the \textit{time step} $T$ dimension. However, its \textit{output channel} $K$ dimension (aka. input channel for next layer input) is completed after the \textit{time step} $T$ dimension, resulting in disorderly sorting of the next layer's input data and inhibiting input sequential addressing. Therefore, we will reorder the temporal channel order as shown in Fig.~\ref{temporal channel reorder fig} by storing the output as the desired input order with non-consecutive addressing. Thus, regardless of the encoding layer or other layers, the output channel dimensions will be arranged sequentially for sequential input data access. The input for the encoding layer will have extra \textit{bit number} $B$ dimension, which will be split and arranged sequentially as shown in the figure.



\subsection{Configuration Setup}
The proposed accelerator has a System Controller as shown in Fig.~\ref{od system architecture fig}, which consists of several registers to store the configurations. The configurations include (1) parameters of convolution to support up to 512 input channels, 512 output channels, and kernel sizes from 1$\times$1 to 3$\times$3, (2) parameters of data flow to  support up to 4 input time steps, 4 output time steps, and 1024$\times$576 input size, (3) number of the sparse weights, (4) two bits for indicating the max-pooling and encoding layer, respectively, and (5) a setup indicator to indicate the setup completion.


\section{Experimental Result}
\subsection{Results of The Proposed Model}
The proposed model is implemented with Pytorch and evaluated  for object detection in the autonomous driving application using the IVS 3cls~\cite{tsai20202020} dataset. IVS 3cls is a cityscape object detection dataset  with about 11,000 samples, including 10,000 training images and 1,000 test images. This dataset contains three types of objects: vehicles, bikes, and pedestrians. Its images have the resolution of 1920$\times$1080. We resize the images to 1024$\times$576 as our input in our experiment. For model training, we use the AdamW~\cite{loshchilov2017decoupled} optimizer with 10$^{-3}$ weight decay. The learning rate is warmed up from 10$^{-5}$ to 10$^{-4}$ at the first five epochs, and finally reduced to 10$^{-6}$. The batch size is set to 32. We train our baseline SNN model for 160 epochs with two NVIDIA Tesla V100.

Table~\ref{ablation study of modified SNN model table} shows the ablation study of the proposed SNN model. SNN-a, the baseline SNN model, requires 3.17M model parameters and achieves mean average precision (mAP) of 73.9$\%$. For model slimming, we fine-tune with an additional 90 epochs for fine-grained pruning, and 5 epochs for quantization and block convolution respectively. The fine-grained pruning causes a 0.6$\%$ mAP drop but reduces about 70$\%$ of parameters. Quantization and block convolution cause 1$\%$ and 0.8$\%$ mAP decrease, respectively. In total, mAP has been dropped by 2.4$\%$, but with 70$\%$ of fewer parameters and much simpler hardware. 


\begin{table*}[htb]
\caption{The ablation study of the SNN model}
\label{ablation study of modified SNN model table}
\centering
\resizebox{0.85\textwidth}{!}{
\begin{tabular}{|c|c|c|c|c|c|c|c|c|}
\hline
\multirow{2}{*}{Model} & \multirow{2}{*}{\begin{tabular}[c]{@{}c@{}}Fine-grained\\ pruning\end{tabular}} & \multirow{2}{*}{\begin{tabular}[c]{@{}c@{}}Quantize\\ (8 bits)\end{tabular}} & \multirow{2}{*}{\begin{tabular}[c]{@{}c@{}}Block\\ convolution\end{tabular}} & \multirow{2}{*}{\begin{tabular}[c]{@{}c@{}}Parameter\\ (M)\end{tabular}} & \multicolumn{4}{c|}{AP (\%)}                \\ \cline{6-9} 
                       &                                                                                 &                                                                              &                                                                              &                                                                          & Bike & Vehicle & Pedestrian & Mean          \\ \hline
SNN-a                  &                                                                                 &                                                                              &                                                                              & 3.17                                                                     & 71.1 & 80.0    & 70.7       & 73.9          \\ \hline
SNN-b                  & \checkmark                                                                               &                                                                              &                                                                              & 0.96                                                                     & 70.6 & 79.1    & 70.2       & 73.3          \\ \hline
SNN-c                  & \checkmark                                                                               & \checkmark                                                                            &                                                                              & 0.96                                                                     & 69.8 & 78.3    & 68.7       & 72.3          \\ \hline
SNN-d                  & \checkmark                                                                               & \checkmark                                                                            & \checkmark                                                                            & 0.96                                                                     & 68.4 & 77.7    & 68.4       & \textbf{71.5} \\ \hline
\end{tabular}}
\end{table*}


\begin{table*}[htb]
\caption{Experimental results of the proposed object detection model}
\label{evaluate result of model table}
\centering
\resizebox{\textwidth}{!}{
\begin{tabular}{|c|c|c|c|c|c|c|c|c|c|}
\hline
\multirow{2}{*}{Model} & \multicolumn{2}{c|}{Precision} & \multirow{2}{*}{\begin{tabular}[c]{@{}c@{}}Block\\ convolution\end{tabular}} & \multirow{2}{*}{\begin{tabular}[c]{@{}c@{}}Model size\\ (Mbits)\end{tabular}} & \multirow{2}{*}{\begin{tabular}[c]{@{}c@{}}Parameter\\ (M)\end{tabular}} & \multicolumn{4}{c|}{AP (\%)}                \\ \cline{2-3} \cline{7-10} 
                       & Act.           & Weight        &                                                                              &                                                                               &                                                                          & Bike & Vehicle & Pedestrian & Mean          \\ \hline
ANN                    & Float32        & Float 32      &                                                                              & 101.44                                                                        & 3.17                                                                     & 79.9 & 81.4    & 79.9       & 80.4          \\
Yolov2~\cite{redmon2017yolo9000}                 & Float 32       & Float 32      &                                                                              & 1618.24                                                                       & 50.57                                                                    & 71.2 & 80.7    & 76.5       & 76.1          \\ \hline
QNN                    & FXP 4          & Float 32      &                                                                              & 101.44                                                                        & 3.17                                                                     & 79.4 & 81.2    & 79.4       & 80.0          \\
QNN                    & FXP 3          & Float 32      &                                                                              & 101.44                                                                        & 3.17                                                                     & 70.6 & 80.3    & 77.3       & 76.1          \\
QNN                    & FXP 2          & Float 32      &                                                                              & 101.44                                                                        & 3.17                                                                     & 69.6 & 78.6    & 67.9       & 72.0          \\
GUO et al.~\cite{guo2020hybrid}             & Hybrid         & Hybrid        &                                                                              & 17.2                                                                          & 6                                                                        & -    & -       & -          & 71.1          \\ \hline
BNN                    & Binary         & Binary        &                                                                              & 3.17                                                                          & 3.17                                                                     & 56.6 & 61.1    & 49.7       & 55.8          \\ \hline
SNN-a                  & Binary         & Float 32      &                                                                              & 101.44                                                                        & 3.17                                                                     & 71.1 & 80.0    & 70.7       & \textbf{73.9} \\
SNN-4T                 & Binary         & Float 32      &                                                                              & 101.44                                                                        & 3.17                                                                     & 71.3 & 80.0    & 70.9       & \textbf{74.1} \\
SNN-d                  & Binary         & FXP 8         & \checkmark                                                                            & \textbf{7.68}                                                                 & 0.96                                                                     & 68.4 & 77.7    & 68.4       & \textbf{71.5} \\ \hline
\end{tabular}}
\end{table*}

For comparison purpose, we also train several models of the same structure with floating point ANN, quantized ANN (QNN), and binary neural network (BNN) as shown in Table~\ref{evaluate result of model table}. The mAP of the baseline SNN-a model is slightly lower than the mAP of the QNN model with three bits activation precision, but much higher than the mAP of the BNN model, as shown in the table. SNN's multiple time steps allow it to extract more features with greater precision. To further evaluate the inference accuracy with more time steps, we take the model weights of SNN-a but infer the model with the (1, 4) mixed time steps (denoted as the SNN-4T model). The result shows that SNN-4T can slightly improve the mAP. In addition, we also compare our SNN-d model with the hybrid precision ANN model ~\cite{guo2020hybrid} trained on the same dataset. The SNN-d has a higher mAP but requires smaller model size and is hardware friendly, demonstrating SNN's effectiveness with small time steps.

\begin{figure}
\begin{subfigure}{1.0\linewidth}
\centering
\includegraphics[height=!,width=1.0\linewidth,keepaspectratio=true]
{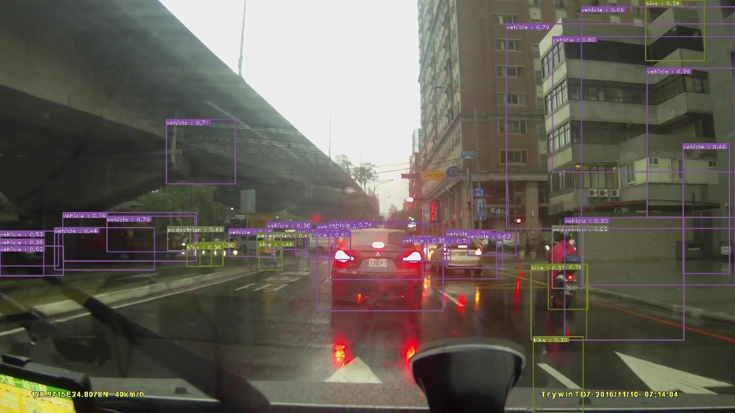}
\caption{Visualization result of the SNN-d model with one time step}
\label{Visualization result of modified SNN model at first time step fig}
\end{subfigure}
\begin{subfigure}{1.0\linewidth}
\centering
\includegraphics[height=!,width=1.0\linewidth,keepaspectratio=true]
{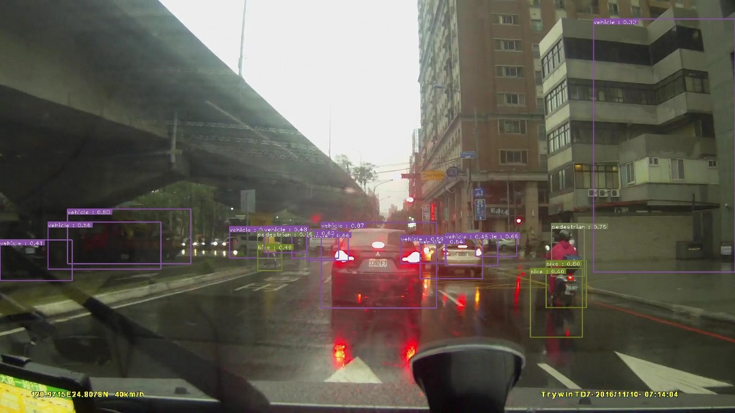}
\caption{Visualization result of the SNN-d model with (1, 2) mixed time steps}
\label{Visualization result of modified SNN model at second time step fig}
\end{subfigure}
\newline
\begin{subfigure}{1.0\linewidth}
\centering
\includegraphics[height=!,width=1.0\linewidth,keepaspectratio=true]
{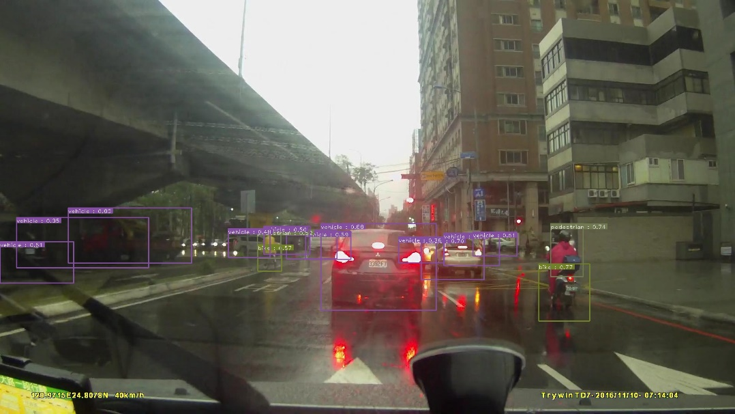}
\caption{Visualization result of the SNN-d model with (1, 3) mixed time steps}
\label{Visualization result of modified SNN model at third time step fig}
\end{subfigure}
\begin{subfigure}{1.0\linewidth}
\centering
\includegraphics[height=!,width=1.0\linewidth,keepaspectratio=true]
{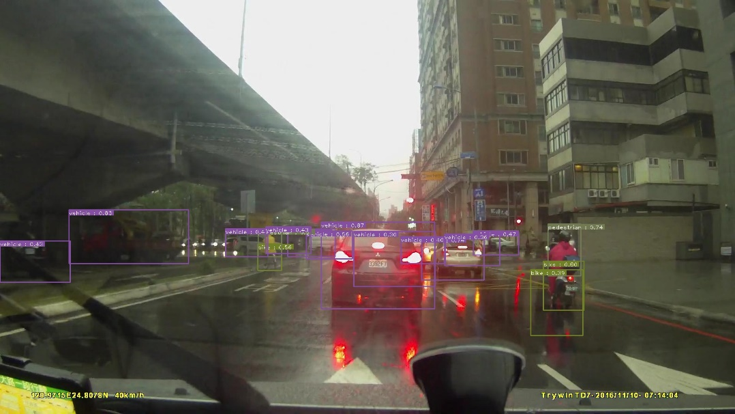}
\caption{Visualization result of the SNN-d model with (1, 4) mixed time steps}
\label{Visualization result of modified SNN model at fourth time step fig}
\end{subfigure}
\caption{Visualization result of the SNN-d model under different time steps}
\label{Visualization result of modified SNN model at time step fig}
\end{figure}

Fig.~\ref{Visualization result of modified SNN model at time step fig} shows the visualization results of the SNN-d model  at various time steps to detect vehicles, bikes, and pedestrians. With only one time step, the result has many false bounding boxes. This situation improves as more time steps are added, as shown in Fig. ~\ref{Visualization result of modified SNN model at time step fig}(b) and (c). The results with (1, 3) and (1, 4) time steps are nearly identical. More time steps can result in more accurate results. Based on this observation, we chose (1, 3) mixed time steps to train our model in order to strike a balance between accuracy and latency.

\subsection{Analysis of Mixed Time Steps}
\begin{figure}[htb]
\centering
\includegraphics[height=!,width=1.0\linewidth,keepaspectratio=true]{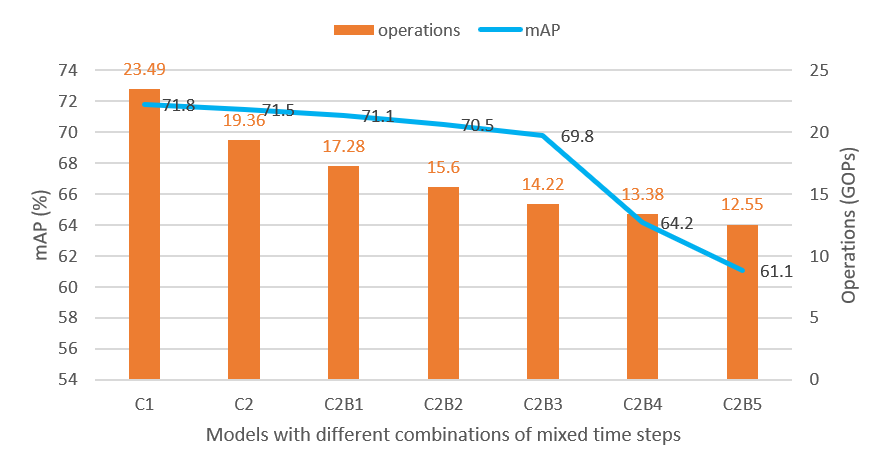}
\caption{The effect of mixed time steps on accuracy and operation count. }
\label{mAP vs operations fig}
\end{figure}

To demonstrate the relationship between the metric mIoUT and the accuracy/operation count, we infer the trained model with different combinations of mixed time steps as shown in Fig.~\ref{mAP vs operations fig}. The C1 model in this diagram denotes that just the first convolutional layer takes one-time-step input and produces three-time-step outputs to the next layer. The C2 model specifies that the first two convolutional layers receive one-time-step input, whereas the second convolutional layer creates three-time-step outputs for the next layer. The C2BX model states that the first X basic blocks, as well as the first two convolutional layers, get one-time-step input, and the basic block's 1$\times$1 convolutional layer creates three-time-step outputs to the following layer. The results show that setting the time step of the first few layers with high mIoUT to 1 can greatly reduce operations while maintaining high accuracy. However, if the time step of the last few layers with low mIoUT is set to 1, the accuracy drops significantly without much reduction in operation counts. As a result, we selected the C2 model to achieve real-time and accurate object detection.


\subsection{Hardware Implementation Result}
\begin{figure}[htb]
\centering
\includegraphics[height=!,width=1.0\linewidth,keepaspectratio=true]
{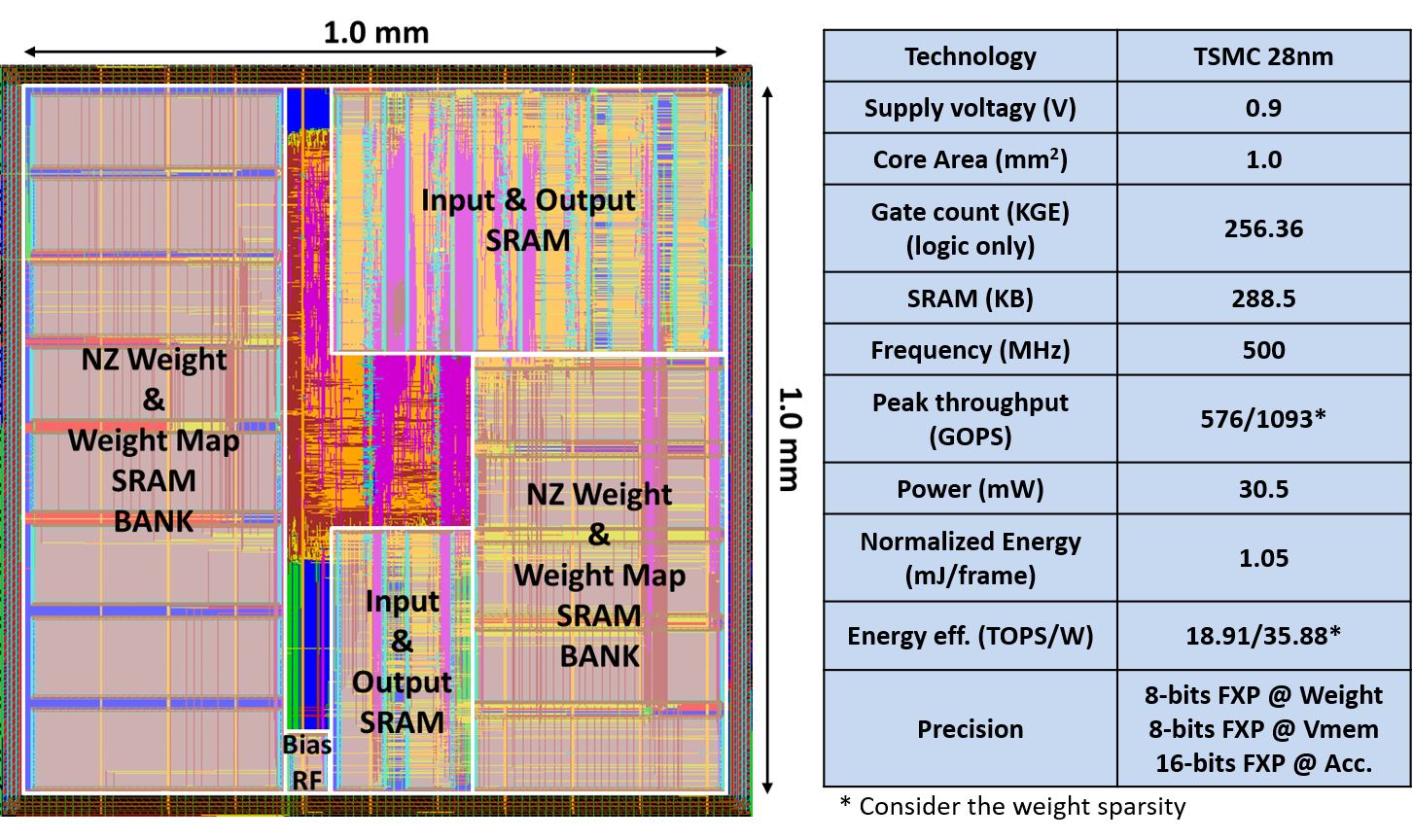}
\caption{Chip layout and its implementation result}
\label{implementation result fig}
\end{figure}

The proposed accelerator has been designed with Verilog, synthesized with Synopsys Design Compiler, and implemented with the TSMC 28nm CMOS technology. Fig.~\ref{implementation result fig} shows the chip layout and its implementation result. The core area is 1.0mm$\times$1.0mm with 288.5KB SRAM. The peak throughput is 1093 GOPS running at 500MHz when considering the weight sparsity. The core power consumption is 30.5mW, measured by running the SNN-d model at 0.9$V$ and 25$^\circ C$. The accelerator can execute our object detection model at 29fps for 1024$\times$576 images and consume only 1.05mJ per frame.

\subsection{External Memory Access Analysis}
To reduce high power external memory access, the size of the NZ Weight SRAM and the Weight Map SRAM are designed to be large enough to store the weight data of the largest layer of the network so that the design will access the local buffer instead of the external DRAM. The size of the Input SRAM is designed to be enough to store a 32$\times$18 tile with 512 input channels and one time step. However, our model runs with (1, 3) mixed time steps. Thus, the inputs of the last few layers will be accessed from the external DRAM repeatedly whenever changing to process different output channels. Therefore, the external access amount for one input image can be divided into three parts: 188.928MB for input, 3.327MB for output, and 1.292MB for model parameters. If we increase the Input SRAM size to be large enough (81KB) to store a 32$\times$18 tile with 384 input channels and three time steps, we can reduce the input access amount to 5.456 MB.

Based on the above analysis, assuming that the external DRAM energy is $70pJ/bit$ for DDR3 DRAM~\cite{malladi2012towards}, the total DRAM access of this design consumes $108.38mJ$ per frame when the Input SRAM is 36KB and $5.64mJ$ per frame when the input SRAM is increased to 81KB. In contrast, the core consumes 1.05mJ per frame. 

\begin{figure}[t]
\centering
\includegraphics[height=!,width=1.0\linewidth,keepaspectratio=true]
{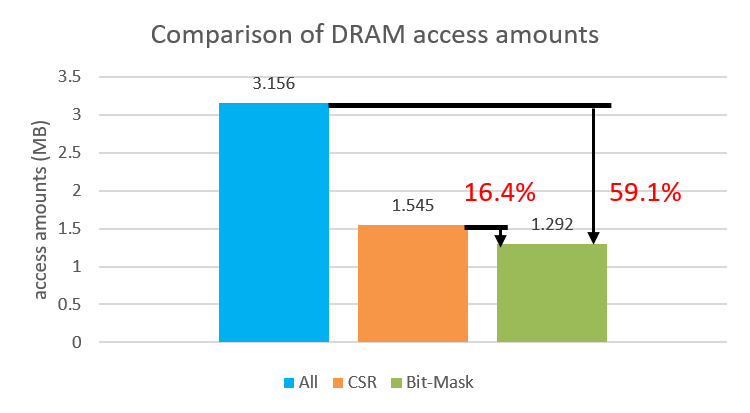}
\caption{The comparison of DRAM access amounts of the network parameters with different representations}
\label{comparison of DRAM access amounts with different representation fig}
\end{figure}

 Fig.~\ref{comparison of DRAM access amounts with different representation fig} compares the DRAM access amounts of network parameters with different representations. Our simple bit-mask representation can reduce 59.1$\%$ and 16.4$\%$ of access compared with the original format and CSR format, respectively.

\subsection{Analysis of Latency, Power and Area}
To analyze the latency benefit due to the zero weight skipping, we design a dense architecture as the baseline, which turns off the skipping to execute all weights. 
With the help of zero weight skipping, we can save 47.3$\%$ of computing latency and execute the model at 29fps for 1024$\times$576 images. The data bandwidth is 5.6GB/s, which is falling within the range of DDR3 DRAM bandwidth (12.8GB/s).

We simulate the design with a smaller dataset, which contains different patches of different layers, and use Synopsys PrimeTime PX to estimate the power consumption. First, with zero activation gating, 
we can decrease 46.6$\%$ of PE dynamic power. The corresponding average sparsity of input maps is 77.4$\%$ without counting the multibit inputs of the first layer.

\begin{figure*}[htb]
\centering
\includegraphics[height=!,width=1.05\linewidth,keepaspectratio=true]
{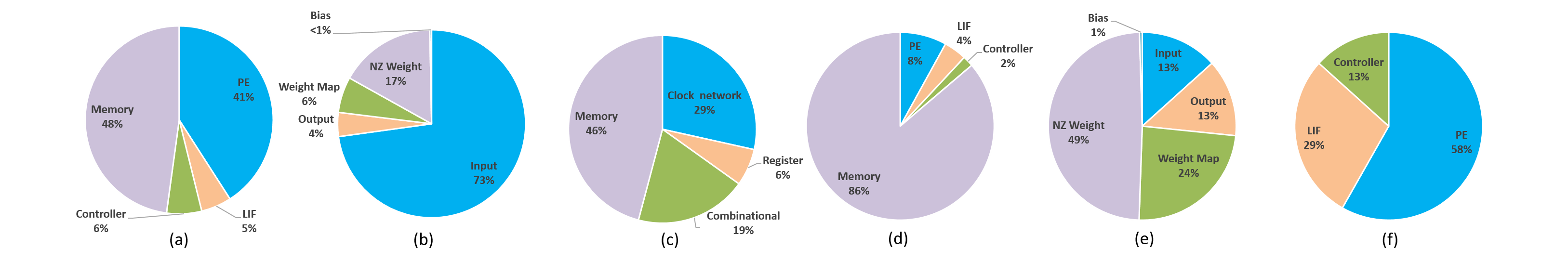}
\caption{Power breakdown of the (a) core, (b) memory and (c) logic circuits. Area breakdown of the (d) core with memory, (e) memory, and (f) core only.}
\label{area and power breakdown}
\end{figure*}

Fig.~\ref{area and power breakdown} shows the power and area breakdown of our design. For power, memory and PEs account for 48$\%$ and 41$\%$ of total power, respectively, because 576 calculation elements need to access large amounts of data and accumulate the partial sum. The input memory accounts for up to 73$\%$ of memory power because of the 144 bit width and simultaneous access of four input SRAMs whenever the input channel is changed. The clock network consumes the 29$\%$ of total power because there are many registers in PE and LIF for accumulating data. For area, the memory occupies 86$\%$ of the total area, and the rest only occupies 14$\%$. PEs occupy 58$\%$ of the logic area due to 576 16-bit registers to accumulate the partial sum. NZ Weight and Weight Map occupy 49$\%$ and 24$\%$ of the total area because they must be large enough to store an entire layer of weights and thus avoid repeatedly accessing the same data from external memory.

\subsection{Design Comparison}
Table~\ref{Comparison with other designs table} shows comparisons with other designs. The list of comparison is selected based on the similarity of the approach to our design. The proposed design achieves area efficiency of 1093 $GOPS/mm^{2}$ and energy efficiency of 35.88 $TOPS/W$ that exceeds most of the other designs due to our simple calculation elements and simple control for skipping zero weights. The area efficiency of \cite{chen202167} is better than ours because their input size is only 80$\times$80, which is much smaller than our 1024$\times$576 input. Thus, the requirements of the I/O memory size are low.

\begin{table}[!htb]
\centering
\caption{Comparison with other designs}
\label{Comparison with other designs table}
\begin{tabular}{|l|l|l|l|l|}
\hline
                                                                  & \textbf{Our Work}                                            & \cite{chen202167}                                                 & \cite{narayanan2020spinalflow}                                             & \cite{park20197}                                              \\ 
\hline
Technology                                                        & 28nm                                                         & 28nm                                                              & 28nm                                                          & 65nm                                                           \\ 
\hline
Measurements                                                      & layout                                                       & layout                                                            & layout                                                        & Chip                                                           \\ 
\hline
Algorithm                                                         & SNN                                                          & SNN                                                               & SNN                                                           & \begin{tabular}[c]{@{}l@{}}SNN \\(MLP)\end{tabular}            \\ 
\hline
Task                                                              & \begin{tabular}[c]{@{}l@{}}Object\\ Detection\end{tabular}   & \begin{tabular}[c]{@{}l@{}}Semantic\\ Segment.\end{tabular}       & CLS                                                           & \begin{tabular}[c]{@{}l@{}}CLS +\\learning\end{tabular}        \\ 
\hline
Sparse                                                            & Y                                                            & Y                                                                 & Y                                                             & N                                                              \\ 
\hline
Voltage (V)                                                       & 0.9                                                          & 0.81                                                              & -                                                             & 0.8                                                            \\ 
\hline
Weight (bits)                                                     & 8                                                            & 8                                                                 & 8                                                             & 8                                                              \\ 
\hline
\# of MAC                                                         & 576 (adder)                                                  & -                                                                 & 128 (adder)                                                   & -                                                              \\ 
\hline
Clock(MHz)                                                        & 500                                                          & 500                                                               & 200                                                           & 20                                                             \\ 
\hline
Peak GOPS$^{a}$                                              & \begin{tabular}[c]{@{}l@{}}576\\1093$^{b, c}$\end{tabular} & \begin{tabular}[c]{@{}l@{}}1150\\1150$^{b}$\end{tabular}     & \begin{tabular}[c]{@{}l@{}}51.2\\51.2$^{b}$\end{tabular} & -                                                              \\ 
\hline
\begin{tabular}[c]{@{}l@{}}Core Area \\(mm$^2$)\end{tabular}      & 1.0                                                          & 0.89                                                              & 2.09                                                          & 10.08                                                          \\ 
\hline
SRAM (KB)                                                         & 288.5                                                        & 240                                                               & 585                                                           & 353                                                            \\ 
\hline
\begin{tabular}[c]{@{}l@{}}Area eff. \\(GOPS/mm$^2$)\end{tabular} & \begin{tabular}[c]{@{}l@{}}576\\1093$^{b, c}$\end{tabular} & \begin{tabular}[c]{@{}l@{}}1292.1\\1292.1$^{b}$\end{tabular} & \begin{tabular}[c]{@{}l@{}}24.5\\24.5$^{b}$\end{tabular} & -                                                              \\ 
\hline
Core Power (mW)                                                   & 30.5                                                        & 149.3                                                             & 162.4                                                         & 23.6                                                           \\ 
\hline
\begin{tabular}[c]{@{}l@{}}Energy eff. \\(TOPS/W)\end{tabular}    & \begin{tabular}[c]{@{}l@{}}18.9\\35.88$^{c, d}$\end{tabular}                                  & \begin{tabular}[c]{@{}l@{}}7.70\\6.24$^{d}$\end{tabular}    & \begin{tabular}[c]{@{}l@{}}-\\ -\end{tabular}                 & \begin{tabular}[c]{@{}l@{}}3.4\\6.24$^{d}$\end{tabular}  \\ 
\hline
\multicolumn{5}{|l|}{$^{a}$1 GMACS = 2 GOPS.}                                                                                                                                                                                                                                                                                         \\
\multicolumn{5}{|l|}{$^{b}$Technology scaling $\left (\frac{process}{28nm} \right )$.}                                                                                                                                                                                                                                                \\
\multicolumn{5}{|l|}{$^{c}$Consider the weight sparsity.}                                                                                                                                                                                                                                                                             \\
\multicolumn{5}{|l|}{$^{d}$Normalized efficiency = efficiency $\times \left ( \frac{process}{28nm} \right )\times \left ( \frac{voltage}{0.9V} \right )^{2}$.}                                                                                                                                                                        \\
\hline
\end{tabular}
\end{table}

\section{Conclusion}



This paper proposes a sparse compressed spiking neural network accelerator for object detection with the gated one-to-all product. The proposed approach can reduce the dynamic power of the PEs by 46.6$\%$ due to the 77.4$\%$ sparsity of input maps and the computing latency by 47.3$\%$ due to the 70$\%$ sparsity of weights. In addition, the sparse compressed weights enable us to use only 216 KB SRAM banks to store the weights of the largest layer and reduce the DRAM access amounts of parameters by 59.1$\%$. The experimental result of the neural network shows 71.5$\%$ mAP with only (1, 3) mixed time steps on the IVS 3cls dataset.
The implementation result can achieve the detection rate at 1024$\times$576@29 frames per second when operating at 500MHz clock frequency and an energy efficiency of 35.88TOPS/W with the 1.05mJ energy consumption per frame. 

\section*{Acknowledgment}
The authors would like to thank TSRI for its support with
EDA design tools.


\bibliographystyle{IEEEtran}


%
%
\bibliography{IEEEabrv,bib/thesis}






\begin{IEEEbiography}[{\includegraphics[width=1in,height=1.25in,clip,keepaspectratio]{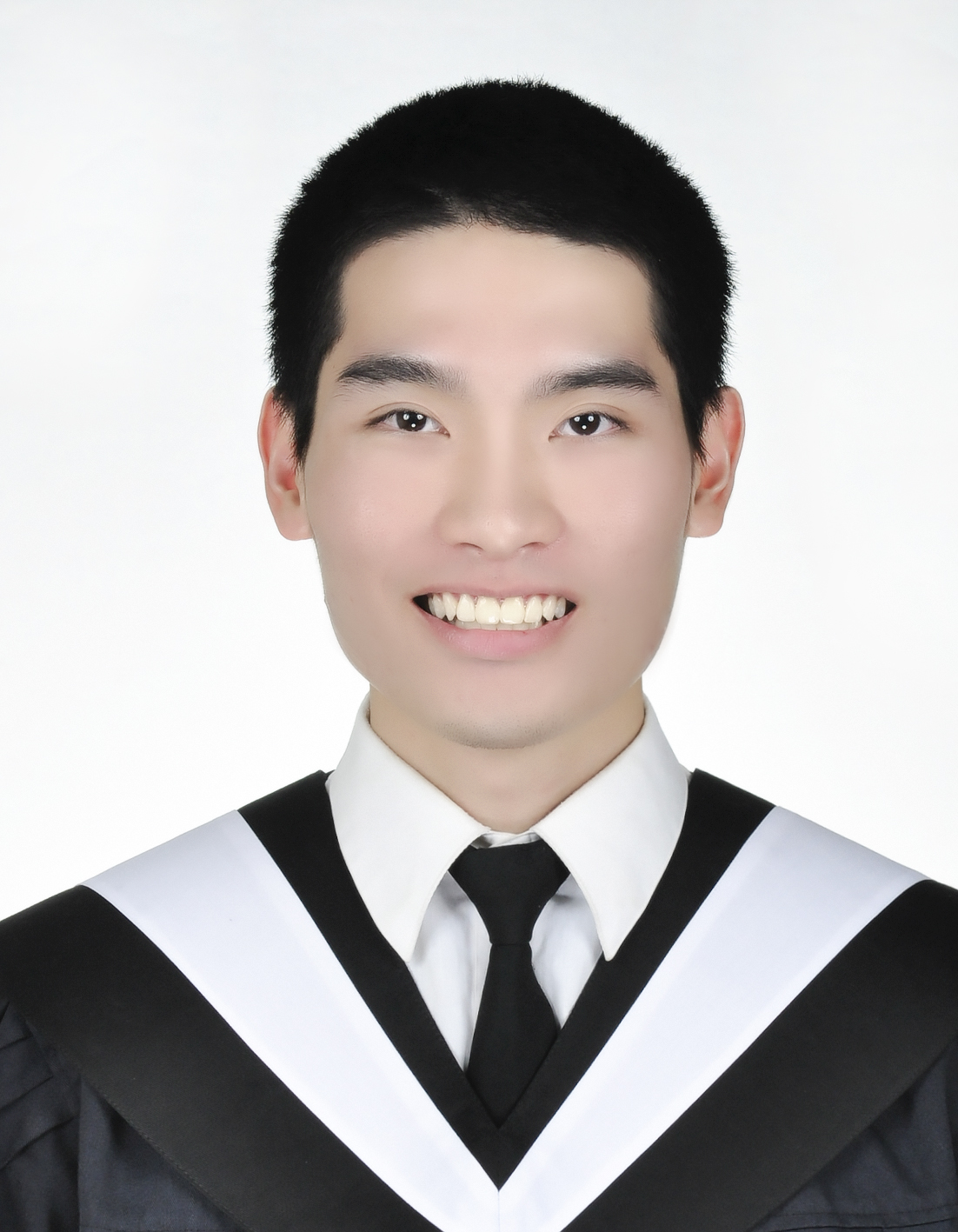}}]{Hong-Han Lien}
received the M.S. degree in the graduate degree program of artificial intelligence from the National Yang Ming Chiao Tung University, Hsinchu, Taiwan, in 2021. He is currently working in the MediaTek, Hsinchu, Taiwan, R.O.C. His research interest includes VLSI design and deep learning.

\end{IEEEbiography}

\begin{IEEEbiography}[{\includegraphics[width=1in,height=1.25in,clip,keepaspectratio]{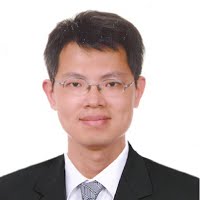}}]{Tian-Sheuan Chang}
	(S’93–M’06–SM’07)
	received the B.S., M.S., and Ph.D. degrees in electronic engineering from National Chiao-Tung University (NCTU), Hsinchu, Taiwan, in 1993, 1995, and 1999, respectively. 
	
	From 2000 to 2004, he was a Deputy Manager with Global Unichip Corporation, Hsinchu, Taiwan. In 2004, he joined the Department of Electronics Engineering, NCTU, where he is currently a Professor. In 2009, he was a visiting scholar in IMEC, Belgium. His current research interests include system-on-a-chip design, VLSI signal processing, and computer architecture.
	
	Dr. Chang has received the Excellent Young Electrical Engineer from Chinese Institute of Electrical Engineering in 2007, and the Outstanding Young Scholar from Taiwan IC Design Society in 2010. He has been actively involved in many international conferences as an organizing committee or technical program committee member.
\end{IEEEbiography}

\end{document}